\documentclass[preprint,5p]{elsarticle}

\usepackage{graphicx}% Include figure files
\usepackage{dcolumn}% Align table columns on decimal point
\usepackage{bm}% bold math
\usepackage[colorlinks,citecolor=blue]{hyperref}
\usepackage[colorinlistoftodos]{todonotes}
\usepackage{ulem}
\usepackage{amsmath,amssymb}

\begin{document}

\title{Non-perturbative collective inertias for fission: a comparative study}

\author[nscl]{Samuel A. Giuliani}
\ead{giuliani@nscl.msu.edu}

\author[ccs,uam]{Luis M. Robledo}
\ead{luis.robledo@uam.es}

\address[nscl]{
  NSCL/FRIB Laboratory,
  Michigan State University, 
  East Lansing, Michigan 48824, USA
}

\address[ccs]{
  Center for Computational Simulation,
  Universidad Polit\'ecnica de Madrid,
  Campus de Montegancedo, 
  Boadilla del Monte, 28660-Madrid, Spain
}

\address[uam]{
  Departamento de F{\'i}sica Te{\'o}rica, 
  Universidad Aut{\'o}noma de Madrid, 
  28049 Madrid, Spain
}%

\date{\today}% It is always \today, today,
             %  but any date may be explicitly specified

\begin{abstract}
The non-perturbative method to compute Adiabatic Time Dependent Hartree Fock Bogoliubov (ATDHFB) 
collective inertias is extended to the Generator Coordinate Method (GCM) in the Gaussian overlap 
approximation (GOA) including the case of density
dependent forces. The two inertias schemes are computed along the fission path of the $^{234}$U 
and compared with the perturbative results. We find that the non-perturbative schemes predict very 
similar collective inertias with a much richer structure than the one predicted by perturbative calculations.
Moreover, the non-perturbative inertias show an extraordinary similitude with the exact GCM inertias computed numerically from the energy overlap. These results indicate that the non-perturbative inertias provide the right structure as a
function of the collective variable and only a phenomenological factor is required to mock up the exact GCM
inertia, bringing new soundness to the microscopic description of fission. 

\end{abstract}

\maketitle

\begin{keyword}

%% keywords here, in the form: keyword \sep keyword
  Fission \sep 
  Collective inertias \sep 
  Generator Coordinate Method \sep
  Adiabatic Time Dependent Hartree Fock Bogolyubov

%% MSC codes here, in the form: \MSC code \sep code
%% or \MSC[2008] code \sep code (2000 is the default)

\end{keyword}

\section{\label{sec:Intro}Introduction}

Despite its discovery dates back almost 80 years, fission still remains a major
challenge for nuclear theory \cite{Schunck2016}. The lack of a feasible
full quantum formalism describing the
evolution of the nucleus from the ground state to scission enforces
the adoption of different approximations, which in turn provide a theoretical
framework for the estimation of fission properties in nuclei. For instance,
the starting point in any traditional energy density functional calculation is
the original assumption that fission can be described using a reduced set of collective
variables \cite{Bohr1939,Hill1953}.  Within this approximation the fission 
probability is obtained as the probability of the nucleus to tunnel under 
the fission barrier, which is driven by the potential 
energy surface (PES) and the collective inertias felt by the nucleus in its 
way to scission \cite{brack1972,bjornholm1980}. Both quantities, together
with the collective ground-state energy, enter in the collective action integral
allowing for the calculation of the spontaneous fission lifetime by means of the
semiclassical Wentzel-Kramers-Brillouin (WKB) approach. 

A sound calculation of the PES, collective ground-state energy and collective inertias 
is thus essential for a proper estimation of fission lifetimes~\cite{Schunck2016,krappe2012}. 
If the formalism that shall be used in the calculation of the first two
quantities is well established, the same cannot be claimed for the collective
inertias. Nowadays two theoretical frameworks allow for a derivation of
a collective Schr{\"o}dinger equation and its associated inertia:
the Adiabatic Time Dependent Hartree Fock Bogoliubov (ATDHFB) formalism  and the
Generator Coordinate Method (GCM) with the Gaussian overlap approximation
(GOA) \cite{Schunck2016}. In both approaches the collective inertias can be written in terms of the
collective momentum operators, which in turn can be related to the 
linear response matrix (LRM) and its inverse when acting on Hartree Fock Bogoliubov (HFB) wave functions. Given the 
whopping number of two-quasiparticle elementary excitations in realistic applications
to fission, the dimensionality of the LRM is very high and therefore
its inverse is difficult to evaluate. To avoid this bottleneck the 
assumption of diagonal dominance of the LRM is often used, leading to the traditional
perturbative cranking formulas for the collective inertias involving denominators composed of 
two quasiparticle energies. A better approximation to the exact expression of
the collective inertias was introduced in~\cite{Baran2011}
(see also~\cite{Yuldashbaeva19991}), where the collective momentum operator 
is computed in terms of the derivatives of the
density and pairing tensor with respect of the collective variables. This
non-perturbative cranking calculation of the collective inertias, implemented in the
ATDHFB approach, showed that the numerical treatment of the derivatives gives
rise to a less adiabatic behaviour of the collective inertias in comparison to
the perturbative calculation.

It follows then that there are two different sources of uncertainty in the
calculation of the collective inertias: one related to the choice of the
theoretical framework (ATDHFB vs GCM-GOA) and the other related to the approximations
involved in the numerical evaluation of the inertias (exact vs non-perturbative
vs perturbative). The purpose of this paper is then twofold:  (i) to introduce
for the first time the non-perturbative scheme in the GCM-GOA framework and (ii)
compute the exact GCM-GOA collective inertias and use this
result to study the suitability of both the ATDHFB  and the GCM-GOA non-perturbative
schemes. Using the actinide $^{234}$U as a test case, we will show that, as the
level of approximation improves, the results obtained in the ATDHFB and the GCM
schemes naturally converge towards the same solution of the collective inertias,
bringing new solidity to the theoretical description of fission. The present
results represent a step forward in the microscopic description of fission 
providing the method with the credibility required to answer questions like
the very existence of nuclei beyond oganesson~\cite{Nazarewicz2018}.

%actinide $^{234}$U is chosen for the numerical applications as a typical example
%in fission applications.

%This paper is organized as follows. In section~\ref{sec:Methods} we firstly
%derive the expression for the momentum operator associated to the collective
%coordinates and we introduce the expression of the collective inertias in the in 
%collective inertias in the perturbative and non-pertubative approximation 

% -----------------------------------------------------------------------------
%                                                   M e t h o d o l o g y 
% -----------------------------------------------------------------------------
\section{\label{sec:Methods}Methodology}

This section is devoted to the derivation of the different expressions used for
the  calculation of the collective inertias. The key element is the momentum
operator $\hat{P}_q$ associated to the collective variable $q$ which is derived
in the quasiparticle representation in section~\ref{subsec:Momentum}. This
quantity is then used to obtain the non-perturbative expression of the GCM-GOA mass
in section~\ref{subsec:GCM} while the extension to density dependent forces is 
presented in section~\ref{subsec:DD}. In section~\ref{subsec:GCM_EX} we discuss how to compute the 
exact GCM-GOA mass using the numerical derivatives of the exact  
Hamiltonian and  norm kernels.
In section~\ref{subsec:ATDHFB} we will briefly
review the derivation of the ATDHFB non-perturbative formula. Section
\ref{subsec:PERT} is devoted to obtaining the explicit expression of the
perturbative masses, both in the GCM-GOA and ATDHFB framework. Finally, the 
connection between collective inertias and moments of inertia is presented in section~\ref{subsec:MOM}.

% -----------------------------------------------------------------------------
%                                                   Linear Momentum operators 
% -----------------------------------------------------------------------------
\subsection{\label{subsec:Momentum}Momentum operator}
Given a collective variable $q$ like, for instance, the quad\-ru\-po\-le moment, its
associated collective moment $\hat{P}_q$ can be defined through the relation (we use $\hbar=1$ in 
the following)
\begin{equation}
i \hat{P}_q |\Phi (q)\rangle = \frac{\partial}{\partial q}|\Phi (q)\rangle = 
\lim_{\delta q\rightarrow 0} \frac{|\Phi (q+\delta q)\rangle - |\Phi (q)\rangle}{\delta q} \,.
\end{equation}
Both $ |\Phi (q+\delta q)\rangle$ and $ |\Phi (q)\rangle$ are HFB wave functions satisfying the HFB equation with the 
corresponding constraints $\langle \Phi (q+\delta q) | \hat{Q}_{20} |\Phi (q+\delta q)\rangle= q+\delta q$ and 
$\langle \Phi (q) | \hat{Q}_{20} |\Phi (q)\rangle= q$, respectively. To evaluate $ |\Phi (q+\delta q)\rangle $ in terms
of $|\Phi (q)\rangle$ we use  linear response theory in the quasiparticle representation.
%, which is not as popular as 
%linear response in the traditional particle representation. For this reason we give some extra details of the 
%derivation below. 
We notice that 
$ |\Phi (q+\delta q)\rangle $  and
$ |\Phi (q)\rangle $ are related by a Thouless transformation that, for infinitesimal $\delta q$, can be written as 
\begin{equation}\label{eq:Phiqdq}
|\Phi (q+\delta q)\rangle = |\Phi (q)\rangle + \delta q \hat{Z} (q) |\Phi (q)\rangle + O(\delta q^2) \,,
\end{equation}
with
\begin{equation}\label{eq:Zq}
\hat{Z} (q) = \frac{1}{2} \sum_{\mu \nu} Z_{\mu \nu} (q) \beta^\dagger_\mu (q) \beta^\dagger_\nu (q) \,.
\end{equation}
On the other hand, $|\Phi (q+\delta q)\rangle $ satisfies the HFB equation with constraints 
\begin{equation}\label{eq:HFBq+dq}
\langle \Phi (q+\delta q) | \left[ \hat{H} -\sum_j \lambda_j (q+\delta q) \hat{Q}_j \right] 
\beta^\dagger_\mu \beta^\dagger_\nu  | \Phi (q+\delta q)\rangle = 0 \,,
\end{equation}
where the quasiparticles creation operators above are defined at deformation $q+\delta q$
\begin{equation}
\beta^\dagger_\mu ( q+\delta q) =\beta^\dagger_\mu ( q) + \delta q \sum_{\mu '} Z^*_{\mu' \mu} \beta_{\mu'}(q) + O(\delta q^2) \,.
\end{equation}
Expanding in powers of $\delta q$ we obtain at zero order 
\begin{equation}
\langle \Phi (q) | \left[ \hat{H} -\sum_j \lambda_j (q) \hat{Q}_j \right] 
\beta^\dagger_\mu (q) \beta^\dagger_\nu (q)  |\Phi (q)\rangle = 0 \,,
\end{equation}
which is an identity because $|\Phi(q)\rangle$ is the constrained HFB solution at deformation $q$. 
At first order in $\delta q$ the following identity has to be satisfied
\begin{equation}\label{eq:R1}
\frac{1}{2}\langle \hat{Z}^\dagger \Delta\hat{H}' \beta^\dagger_\mu \beta^\dagger_\nu \rangle + 
\frac{1}{2}\langle \Delta\hat{H}'   \beta^\dagger_\mu \beta^\dagger_\nu \hat{Z} \rangle =
 \sum_j \frac{\partial \lambda_j}{\partial q} (Q_j)^{20\,*}_{\mu \nu} \,,
\end{equation}
as well as its complex conjugated. In the above expression $\Delta \hat{O} =  \hat{O} - \langle \hat{O} \rangle$,
$(Q_j)^{20\,*}_{\mu \nu}= \langle \beta_\nu \beta_\mu \hat{Q}_j\rangle$ is the 20 part of the operator $\hat{Q}_j$,
and $\hat H'=\hat H -\sum_j \lambda_j \hat Q_j$. Introducing the matrices
\begin{subequations}\label{eq:AB}
\begin{align}
A_{\mu \nu \mu' \nu'} & =   \langle \beta_\nu \beta_\mu \Delta\hat{H}' \beta^\dagger_{\mu'} \beta^\dagger_{\nu'} \rangle \,, \\
B_{\mu \nu \mu' \nu'} & =   \langle \beta_\nu \beta_\mu \beta_{\nu'} \beta_{\mu'}\Delta\hat{H}'  \rangle \,,
\end{align}
\end{subequations}
with the properties $A_{\mu'\nu' \mu\nu}=A_{\mu\nu\mu'\nu'}^*$ and $B_{\mu'\nu' \mu\nu}=B_{\mu'\nu' \mu\nu}$, Eq.~\eqref{eq:R1} becomes
\begin{align}
\sum_{\mu' < \nu'} Z_{\mu' \nu'}^* A_{\mu'\nu'\mu\nu} +
Z_{\mu' \nu'} B^*_{\mu'\nu'\mu\nu} & =  \sum_j \frac{\partial \lambda_j}{\partial{q}} (Q_j^{20})^*_{\mu\nu} \,, \\ \nonumber
\sum_{\mu' < \nu'} Z_{\mu' \nu'}^* B_{\mu'\nu'\mu\nu} +
Z_{\mu' \nu'} A^*_{\mu'\nu'\mu\nu} & =  \sum_j \frac{\partial \lambda_j}{\partial{q}} (Q_j^{20})_{\mu\nu} \,.
\end{align}
To simplify the notation it is convenient to introduce indexes $\rho$ and
$\sigma$ corresponding to the pair of indexes $\mu$ and $\nu$ with the restriction $\mu < \nu$. 
The ordering of the correspondence is irrelevant in what follows. With the new
indexes, $Z_{\mu \nu}$ becomes the vector $Z_\rho$ and the four index quantities
$A_{\mu'\nu'\mu\nu}$ become the matrix elements of a hermitian matrix $A_{\rho' \rho}$. The same applies to
the $B_{\mu'\nu'\mu\nu}$  that become the matrix elements of a symmetric matrix $B_{\rho' \rho}$. 
In terms of the new indexes the previous equation becomes
\begin{subequations}
\begin{align}
\sum_{\rho'} Z_{\rho'}^* A_{\rho'\rho} +
Z_{\rho'} B^*_{\rho'\rho} & =  \sum_j \frac{\partial \lambda_j}{\partial{q}} (Q_j^{20})^*_{\rho} \,, \\
\sum_{\rho'} Z_{\rho'}^* B_{\rho'\rho} +
Z_{\rho'} A^*_{\rho'\rho} & =  \sum_j \frac{\partial \lambda_j}{\partial{q}} (Q_j^{20})_{\rho} \,.
\end{align}
\end{subequations}
Introducing the linear response matrix (LRM) $\mathbb{L}$
\begin{equation}
\mathbb{L} =  \left( \begin{array}{cc} A & B \\ B^* & A^* \end{array}\right),
\end{equation}
which is closely related to the matrix appearing in the Random Phase Approximation (RPA),
it is easy to express $Z$ in terms of the partial derivatives of the chemical potentials
\begin{equation}
\left( \begin{array}{c} Z \\ Z^* \end{array}\right) = \mathbb{L}^{-1}
\sum_j \frac{\partial \lambda_j}{\partial{q}} \left( \begin{array}{c} Q_j^{20} \\ Q_j^{20\,*} \end{array}\right) \,.
\end{equation}
The partial derivatives of the chemical potentials are determined by plugging the above 
result in the definition of the constraints
\begin{equation}
\langle \Phi (q_j+\delta q) | \hat{Q}_{i} |\Phi (q_j+\delta q)\rangle= q_i + \delta_{ij} \delta q \,.
\end{equation}
As a consequence of this requirement we get
\begin{equation}
\frac{\partial \lambda_i}{\partial{q_j}} = \left( M_{(-1)}^{-1} \right)_{ij} \,,
\end{equation}
where the quantity $\left( M_{(-1)} \right)_{ij} $ is given by
\begin{equation}\label{eq:MM1}
\left( M_{(-1)} \right)_{ij} = 
( Q_i^{20\, *} Q_i^{20}) \mathbb{L}^{-1} \left( \begin{array}{c} Q_j^{20} \\ Q_j^{20\,*} \end{array}\right) \,.
\end{equation}
Collecting together  all the partial results we finally obtain
\begin{equation}\label{eq:Pq}
\left( \begin{array}{c} P_{q_j}^{20} \\ -P_{q_j}^{20\,*} \end{array}\right) = -i
\sum_k \left( M_{(-1)}^{-1} \right)_{kj} \mathbb{L}^{-1}   \left( \begin{array}{c} Q_k^{20} \\ Q_k^{20\,*} \end{array}\right) \,.
\end{equation}
The evaluation of the momentum matrix elements requires the inversion of 
$\mathbb{L}$ which is in general a tremendous task, given the typical number of two quasiparticle
excitations involved in a realistic calculation. An alternative (and useful) expression for the
momentum operator (or $Z$) can be obtained by evaluating the derivatives of the densities (both normal
and abnormal) with respect to the constraints (see Appendix).
% -----------------------------------------------------------------------------------
%                                             The GCM inertia
% -----------------------------------------------------------------------------------
\subsection{\label{subsec:GCM}The Generator Coordinate Method inertia}
% -----------------------------------------------------------------------------------
%The GCM by itself does not require an inertia. It is only after considering some
%kind of local approximation that one ends up with a collective Schrodinger equation
%and an inertia \cite{ring2000}. Traditionally, the GOA is the approximation of choice
%to obtain those quantities
The GCM does not directly provide an expression of the collective inertia. 
It is only after introducing some local approximation that the Hill-Wheeler 
equation can be reduced to a collective Schr{\"o}dinger equation and yield the associated
inertia \cite{ring2000}. Traditionally, the GOA is the approximation of choice
to make this connection.
%between the Hill-Wheeler and the collective Schr{\"o}dinger equations. 

Assuming that the width of the Gaussian does not depend on $q$
the GCM-GOA mass is given by \cite{Schunck2016,krappe2012,ring2000}
\begin{equation}\label{eq:MGCM}
  \frac{1}{M_{\mathrm{GOA}}}=-\frac{1}{4\gamma^{2}}(h_{qq}+h_{q'q'}-2h_{qq'}) \,,
\end{equation}
where $h_{qq}=\frac{\partial^{2}}{\partial q^{2}}h(q,q')_{|q=q'}$,
$h_{q'q'}=\frac{\partial^{2}}{\partial q'^{2}}h(q,q')_{|q=q'}$ and
$h_{qq'}=\frac{\partial^{2}}{\partial q\partial q'}h(q,q')_{|q=q'}$
with
\begin{equation} \label{eq:hqq}
   h(q,q')=\frac{\langle\phi(q)|\hat{H}_{\mathrm{eff}}|\phi(q')\rangle}{\langle\phi(q)|\phi(q')\rangle} \,.
\end{equation}
Here $\hat{H}_{\mathrm{eff}} = \hat{H} - \lambda_N (\hat{N}-N) - \lambda_Z (\hat{Z}-Z)$
as required to preserve particle number on the average also for the GCM wave functions
\cite{Gozdz1985a,Bonche1990}.
If the constant width is not assumed \cite{Gozdz1985,Gozdz1985a,Bonche1990} 
the above expression remains valid, but one has to replace the partial derivatives by covariant
ones that include in their definition the affine connection or Christoffel symbols
of differential geometry. We will use in the following the constant width formula
to preserve the traditional connection with the momentum operator defined above. 
Assuming time reversal invariant states $|\phi(q)\rangle$ such that
$\langle\phi(q)|\frac{{\partial}}{\partial q'}|\phi(q')\rangle_{|q'=q}=0$
and computing second derivatives of the HFB states as
$$
\frac{\partial^{2}}{\partial q^{2}}|\phi(q)\rangle=\lim_{\delta q\rightarrow0}\frac{1}{\delta q^{2}}(|\phi(q+\delta q)\rangle+|\phi(q-\delta q)\rangle-2|\phi(q)\rangle) \,,
$$
with 
$|\phi(q+\delta q)\rangle=\mathcal{N}(q)(1+\delta q\hat{Z}+\frac{1}{2}\delta q^{2}\hat{Z}^{2}+\cdots)|\phi(q) \rangle$ (curvature terms $\frac{\partial Z}{\partial q}$ \cite{Reinhard1984} are omitted) we finally obtain 
\begin{align*}
h_{qq} & =\langle\phi(q)|\left(\hat{Z}^{\dagger}\right)^{2}\Delta\hat{H}_\mathrm{eff}|\phi(q)\rangle=
%\sum_{\mu<\nu}\sum_{\mu'<\nu'}Z_{\mu\nu}^{*}Z_{\mu'\nu'}^{*}B_{\mu\nu\mu'\nu'}=
\sum_{\rho\rho'}Z_{\rho}^{*}Z_{\rho'}^{*}B_{\rho\rho'} \,, \\
h_{q'q'} & =\langle\phi(q)|\Delta\hat{H}_\mathrm{eff}\hat{Z}^{2}|\phi(q)\rangle=
%\sum_{\mu<\nu}\sum_{\mu'<\nu'}Z_{\mu\nu}Z_{\mu'\nu'}B_{\mu\nu\mu'\nu'}^{*}=
\sum_{\rho\rho'}Z_{\rho}Z_{\rho'}B_{\rho\rho'}^{*} \,, \\
h_{qq'} & =\langle\phi(q)|\hat{Z}^{\dagger}\Delta\hat{H}_\mathrm{eff}\hat{Z}|\phi(q)\rangle=
%\sum_{\mu<\nu}\sum_{\mu'<\nu'}Z_{\mu\nu}^{*}Z_{\mu'\nu'}A_{\mu\nu\mu'\nu'}=
\sum_{\rho\rho'}Z_{\rho}^{*}Z_{\rho'}A_{\rho\rho'} \,,
\end{align*}
that leads to the compact expression
\begin{equation}
\frac{1}{M_{\mathrm{GOA}}}=\frac{1}{4\gamma^{2}}\left(\begin{array}{cc}
Z^{*} & Z\end{array}\right)\left(\begin{array}{cc}
A & -B \\
-B^{*} & A^{*}
\end{array}\right)\left(\begin{array}{c}
Z \\
Z^{*}
\end{array}\right) \,. \label{eq:MGCM-1}
\end{equation}
Please note that the $A$ and $B$ matrices above are not exactly the
same as those of Eqs.~\eqref{eq:AB} which are defined in terms of $\hat H'$ instead
of $\hat H_\mathrm{eff}$. The differences, associated with the collective constrains,
are zero for the ground state and very small elsewhere as we have checked in our example 
below. In the following we will assume them to be the same. 
Using the definition of $Z$,  $\mathbb{L}$
and introducing the matrix $\eta=\left(\begin{array}{cc}
1 & 0\\
0 & -1
\end{array}\right)$ we finally obtain
$$
\frac{1}{M_{\mathrm{GOA}}}=\frac{1}{4}\gamma^{-1}M_{(-1)}^{-1}\bar{M}_{(-1)}M_{(-1)}^{-1}\gamma^{-1} \,,
$$
that is written in a way that can be easily generalized to the multidimensional
case. The $\bar{M}_{(-1)}$ is given by 
\begin{equation} \label{eq:Mbar}
\bar{M}_{(-1)_{lm}}=\left(\begin{array}{cc}
Q_{l}^{20\,*} & Q_{l}^{20}\end{array}\right)\mathbb{L}^{-1}\eta\mathbb{L}\eta\mathbb{L}^{-1}\left(\begin{array}{c}
Q_{m}^{20}\\
Q_{m}^{20\,*}
\end{array}\right) \,.
\end{equation}
The width $\gamma$ can be obtained in a similar manner: 
\begin{equation}\label{eq:width}
\begin{split}
\gamma &= \frac{\partial^{2}}{\partial q\partial q'}\langle\phi(q)|\phi(q')\rangle_{|q=q'}=
\langle\phi(q)|\hat{Z}^{\dagger}\hat{Z}|\phi(q)\rangle\\
&=\sum_{\rho}|Z_{\rho}|^{2}=\frac{1}{2}M_{(-1)}^{-1}{M}_{(-2)}M_{(-1)}^{-1} \,,
\end{split}
\end{equation}
where ${M}_{(-2)}$ is defined in analogy with Eq.~\eqref{eq:MM1} but replacing 
$\mathbb{L}^{-1}$ by $\mathbb{L}^{-2}$.
In the non-perturbative cranking approach we use in Eq.~\eqref{eq:MGCM-1} the
$Z$ obtained from the partial derivatives of the density matrix and
pairing tensor (see Eq.~\eqref{eq:ZpdR}).
Additionally, we use the cranking approximation for $\mathbb{L}$ where  
$B=0$ and $A$ is replaced by
its diagonal approximation $A_{\rho\rho'}=E_{\rho}^{2qp}\delta_{\rho\rho'}$
with $E_{\rho}^{2qp}=E_{\mu}+E_{\nu}$. Inserting this approximation
into the general equation we arrive to 
\begin{equation}\label{eq:MNPGCM}
\frac{1}{M_{\mathrm{GOA}}^{NP}}=\frac{\sum_{\mu<\nu}(E_{\mu}+E_{\nu})|Z_{\mu\nu}|^{2}}{2\left(\sum_{\mu<\nu}|Z_{\mu\nu}|^{2}\right)^{2}} \,,
\end{equation}
that is the expression used in this paper. We do not use BCS like approximations
like the one discussed in Ref \cite{Baran2011}.

% -----------------------------------------------------------------------------------
%                                      Density dependent forces
% -----------------------------------------------------------------------------------
\subsection{\label{subsec:DD}Density dependent forces}

For density dependent forces like Gogny or Skyrme the above formulation has to
be slightly modified. In Eq.~\eqref{eq:HFBq+dq} the Hamiltonian $\hat H (q+\delta q)$ 
has to be replaced by $\hat H (q+ \delta q) + \partial \hat\Gamma (q+ \delta q)$ 
where the one body rearrangement term is given by 
$\partial \hat\Gamma (q+ \delta q) = \sum_{ij}\partial 
\Gamma (q+\delta q)_{ij} c^\dagger_i c_j$ with matrix elements
\begin{equation}
 \partial \Gamma (q+\delta q)_{ij} =  \langle \Phi (q+\delta q) | \frac{\delta \hat H}{\delta \rho} 
 \varphi^*_i (\vec{r})  \varphi_j (\vec{r})| \Phi (q+\delta q)\rangle \,. 
\end{equation}
When expressing those quantities in terms of the corresponding ones
at deformation $q$, derivatives with respect to $q$ of both $\hat H (q)$
and $\partial \hat\Gamma (q)$  have to be considered. The expressions 
obtained are rather involved but straightforward to derive and they 
will not be given here. In addition,
those derivatives only enter the LRM and therefore they are not required in the non-perturbative case
except for the definition of the one-quasiparticle energies that must be computed 
with the Hamiltonian including
the rearrangement term  $\hat H (q) + \partial \hat\Gamma (q)$.

% -----------------------------------------------------------------------------------
%                                     The exact  Generator Coordinate Method inertia
% -----------------------------------------------------------------------------------
\subsection{\label{subsec:GCM_EX}The GCM-GOA inertia}

To compute the  GCM-GOA inertia without using the cranking approximation
we use Eq.~\eqref{eq:MGCM} evaluating the derivatives numerically. 
The required Hamiltonian overlap in Eq. (\ref{eq:hqq}) is evaluated using 
the expressions of the generalized Wick theorem~\cite{Ber12}. 
For the phenomenological density dependent part of the Gogny force we 
use the mixed density prescription as discussed in Refs~\cite{Rob07,Rob10a}. 
First order finite difference formulas are used for the second derivatives
($f''(x)=(f(x+h)+f(x-h)-2f(x))/h^2$) with a value of $h$ conveniently chosen 
according to the collective variable used (see below). The width $\gamma$ is
computed numerically in the same way from the norm overlap.

% -----------------------------------------------------------------------------------
%                                                The Adiabatic Time Dependent inertia
% -----------------------------------------------------------------------------------
\subsection{\label{subsec:ATDHFB}The Adiabatic Time Dependent HFB inertia}

The ATDHFB inertia~\cite{BARANGER1978123} can be evaluated using the same framework 
as above, but imposing additional  constraints on the momentum operators 
\cite{VILLARS1977269}. We are not going to provide the details here, but
following the same steps as above in the quasiparticle picture one gets
\begin{equation} \label{eq:ATDHFB}
M^\mathrm{ATDHFB}_{lm}=
\left(\begin{array}{cc}P_{l}^{20\,*} & P_{l}^{20}\end{array}\right)
\mathbb{L}^{-1}
\left(\begin{array}{c} P_{m}^{20}\\ P_{m}^{20\,*}\end{array}\right) \,.
\end{equation}
Introducing the matrix
\begin{equation} \label{eq:Mbar3}
\bar{M}_{(-3)_{lm}}=\left(\begin{array}{cc}
Q_{l}^{20\,*} & Q_{l}^{20}\end{array}\right)\mathbb{L}^{-1}\eta\mathbb{L}^{-1}\eta\mathbb{L}^{-1}\left(\begin{array}{c}
Q_{m}^{20}\\
Q_{m}^{20\,*}
\end{array}\right) \,,
\end{equation}
which is very similar in structure to $\bar{M}_{(-1)_{lm}}$ of Eq.~\eqref{eq:Mbar},
we obtain
\begin{equation}
    M^\mathrm{ATDHFB} = M^{-1}_{(-1)} \bar{M}_{(-3)}M^{-1}_{(-1)} \,.
\end{equation}
A method for the exact evaluation of the ATDHFB inertia has been 
formulated in~\cite{DOBACZEWSKI1981123} and applied to very simple cases. 
A more recent attempt based on a direct evaluation of the LRM
and its numerical inversion seems to be rather impractical due 
to the enormous computational cost~\cite{Deloncle16}. Recently, 
it has been suggested~\cite{PhysRevC.92.034321} that the
finite amplitude method~\cite{Nataksukasa14} could be useful for 
this task, but so far, it has only been applied to the evaluation 
of the Thouless-Valatin moment of inertia.

% -----------------------------------------------------------------------------------
%                                      The perturbative (cranking) approximation
% -----------------------------------------------------------------------------------
\subsection{\label{subsec:PERT}The perturbative approximation}

In the perturbative cranking approximation 
the $\mathbb{L}$ matrix is approximated by its diagonal also in
the definition of the momentum operator Eq.~\eqref{eq:Pq}.
The approximate $\mathbb{L}$ commutes 
with $\eta$ and the quantity defined in Eq.~\eqref{eq:Mbar} becomes the 
$M_{(-1)}$ defined in Eq.~\eqref{eq:MM1} which in turn becomes  
$M^\mathrm{PE}_{(-1)}$ with the momenta or order $n$ defined as
\begin{equation}
    \left(M^\mathrm{PE}_{(-n)}\right)_{ij} = 
    \sum_{\mu < \nu } \frac{(Q^{20\,*}_i)_{\mu \nu} (Q^{20}_j)_{\mu \nu} + h.c.}
    {(E_\mu+E_\nu)^n} \,.
\end{equation}
In the perturbative
approximation the $\bar{M}_{(-3)}$ of Eq.~\eqref{eq:Mbar3}
becomes $M^\mathrm{PE}_{(-3)}$ and the $M_{(-2)}$ in the 
definition of the width Eq.~\eqref{eq:width} becomes $M^\mathrm{PE}_{(-2)}$.

% -----------------------------------------------------------------------------------
%                                 Connection with rotational band moments of inertia
% -----------------------------------------------------------------------------------

\subsection{\label{subsec:MOM}Connection with rotational band moments of inertia and 
inequalities}

We would like to mention the similarity between  the non-perturbative 
inertias and the Inglis-Belyavev and approximate Yoccoz moments of
inertia~\cite{ring2000}. In this case, the ``momentum operator'' is dictated by 
symmetry considerations (it is the $J_x$ operator) and therefore the 
approximate expressions used in the literature to compute moments of 
inertia fall into the non-perturbative cranking category 
discussed here~\cite{ring2000}. 
Concerning the exact moments of inertia, the Thouless-Valatin moment of
inertia is obtained in a similar framework to the ATDHFB case, whereas the Yoccoz
moment of inertia corresponds to the GCM-GOA inertia. A comparison of the two 
moments of inertia~\cite{Reinhard1984,RoG00c} reveals that the Thouless-Valatin 
is typically a factor 1.4 larger than the Yoccoz one both when computed exactly 
and when computed in the non-perturbative cranking spirit. 

In~\cite{FIOLHAIS1983205} it was shown, using the Schwarz inequality, that at the
minima of the potential energy surface the exact inertias satisfy $ M^\mathrm{ATDHFB}\ge M^\mathrm{GCM}$. 
The same inequality was proved  true for the whole fission path in the
case of the perturbative inertias. In the case of the non-perturbative inertias
and following the same arguments as in~\cite{FIOLHAIS1983205} it is straightforward
to show that the same inequality also applies.

% -----------------------------------------------------------------------------------
%                                                              F i g u r e   1
% -----------------------------------------------------------------------------------
\begin{figure}
\includegraphics[width=\columnwidth]{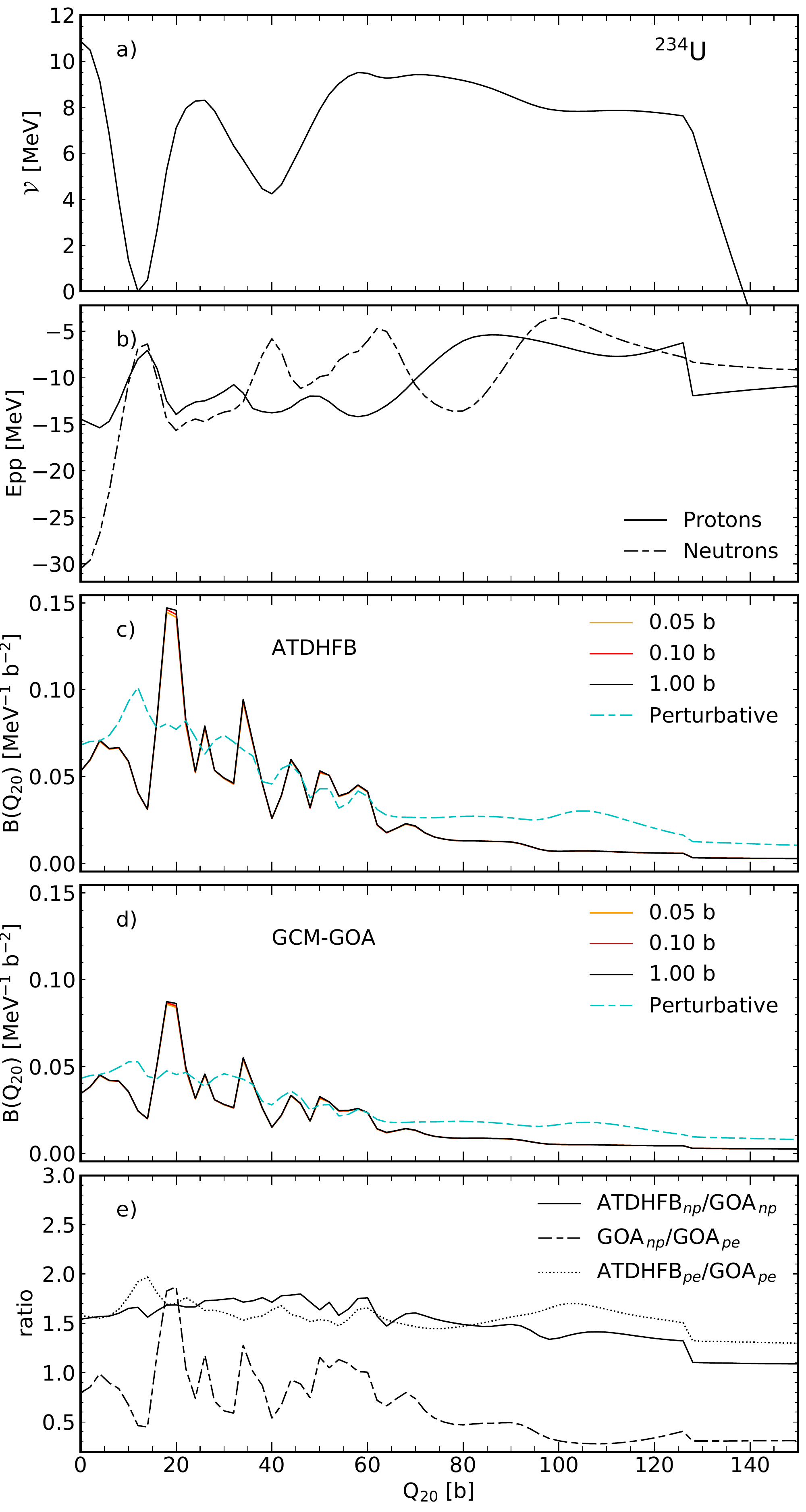}
\caption{\label{fig-1}(Color online) Fission results for the nucleus $^{234}$U
shown as a function of $Q_{20}$ (b). In panel a) the HFB energy is given. The
particle-particle proton (full) and neutron (dashed) energies are plotted. In
panels c) and d) the ATDHFB and GCM-GOA inertias, respectively, are shown in the
perturbative (blue dashed) and non-perturbative (full lines) approaches. In the
non-perturbative case, we observe several overlaid curves corresponding to
different values of a parameter used in the numeric evaluation of the momentum
operator. In panel e) the ratios ATDHFB to GCM-GOA inertias (NP, full line, PE dotted
line) and GCM-GOA$_{np}$ to GCM-GOA$_{pe}$ are shown.}
\end{figure}
% -----------------------------------------------------------------------------------

% -----------------------------------------------------------------------------------
%                                                              An example
% -----------------------------------------------------------------------------------

\section{Results}

In this section we compare the numerical results obtained for the quadrupole
collective inertia in the typical case of the fission of the actinide $^{234}$U. 

%\subsection{Normal deformed nucleus}
%In this section we discuss the results for the inertias in the nucleus
%$^{154}$Sm as a typical example of a well quadrupole deformed nucleus with
%some regions of octupole instability. The potential energy surface and the
%associated $|\phi(q_2)\rangle$ HFB intrinsic states has been generated 

%\subsection{Fissioning nuclei}
%In this section we discuss the results for the different kinds of inertias 
%computed in this paper for the actinide
%nucleus $^{234}$U.

In Fig \ref{fig-1} the results for $^{234}$U are shown as a function of the mass
quadrupole moment $Q_{20}$ expressed in barns. In panel a) the potential energy
surface, given by the HFB energy, is depicted: the characteristic normal deformed
minimum at $Q_{20}=12\,$b ($\beta_2=0.25$) is obtained, followed by a fission
isomer (or super-deformed state) at $Q_{20}=40\,$b ($\beta_2=0.71$) with an
excitation energy of 4.2 MeV. A very broad and high (9.5 MeV) second fission
barrier is found next. In panel b) the particle-particle energy $E_\mathrm{pp}$
defined as $\frac{1}{2} \mathrm{tr} (\Delta^*\kappa)$ is given separately for
protons (full) and neutrons (dashed). A rather intricate behaviour is observed
with the quadrupole moment, due to different level crossings that increase or
decrease the level density around the Fermi level. The collective inertias
[ATDHFB, panel c)] and  [GCM-GOA, panel d)] are also given in the figure. The full
and dashed lines represent the non-perturbative (NP) and perturbative (PE)
cranking results, respectively. Full lines labeled with the values 0.05 b, 
0.1 b and 1 b
represent calculations with different values of the step size $\delta Q_{20}$
used to evaluate the derivatives numerically. The different curves sit one on
top of each other indicating a satisfactory convergence of the method.  

Two main conclusions regarding the collective inertias can be
extracted from panels c) and d). The first one is that the ATDHFB$_{np}$ and
GCM-GOA$_{np}$ cranking inertias have exactly the same peak structure. 
As expected (see 
the discussion in Sec. \ref{subsec:MOM}), and in both the NP and PE cases,  the
ATDHFB mass is  larger than the GCM-GOA one, as shown in panel e). 
What is remarkable is that in the two numerical schemes (NP and PE) the ATDHFB 
and GCM-GOA 
inertias differ by a 
factor around 1.5 and rather constant over the whole deformation range
up to the region corresponding to two separate fragments ($Q_{20} > 126\,$b). After
this point on the quadrupole moment has a geometric origin (being proportional to
the square of the distance between the two fragments) and the
ATDHFB$_{np}$/GCM-GOA$_{np}$ ratio remains very close to one.  We also point out
that the 1.5 value of the ratio is consistent with other studies concerning the
values of the Thouless-Valatin and Yoccoz moments of inertia (see
section~\ref{subsec:MOM}).  The
second finding is that both the ATDHFB$_{np}$ and GCM-GOA$_{np}$ inertias have more
pronounced peaks compared to the perturbative calculations. Looking at the
variations of the particle-particle energy $E_{pp}$ it is possible to relate
these peaks with a larger sensitivity of the non-perturbative inertias to the
presence of level crossings. This ``lack of adiabaticity''  of the non-perturbative
inertias is consistent with the ATDHFB results of Baran \textit{et
al.}~\cite{Baran2011}.
%and here we show for the first time the extension to the GCM scheme. 
Finally, the fluctuations of the GCM-GOA$_{np}$/GCM-GOA$_{pe}$ ratio depicted in
panel e) (dashed line) suggest that the recipe of multiplying the perturbative
inertia by a constant factor in order to simulate the non-perturbative masses is
not a reasonable assumption~\cite{Libert10}.

% -----------------------------------------------------------------------------------
%                                                              F i g u r e   2
\begin{figure}[tb]
\includegraphics[width=\columnwidth]{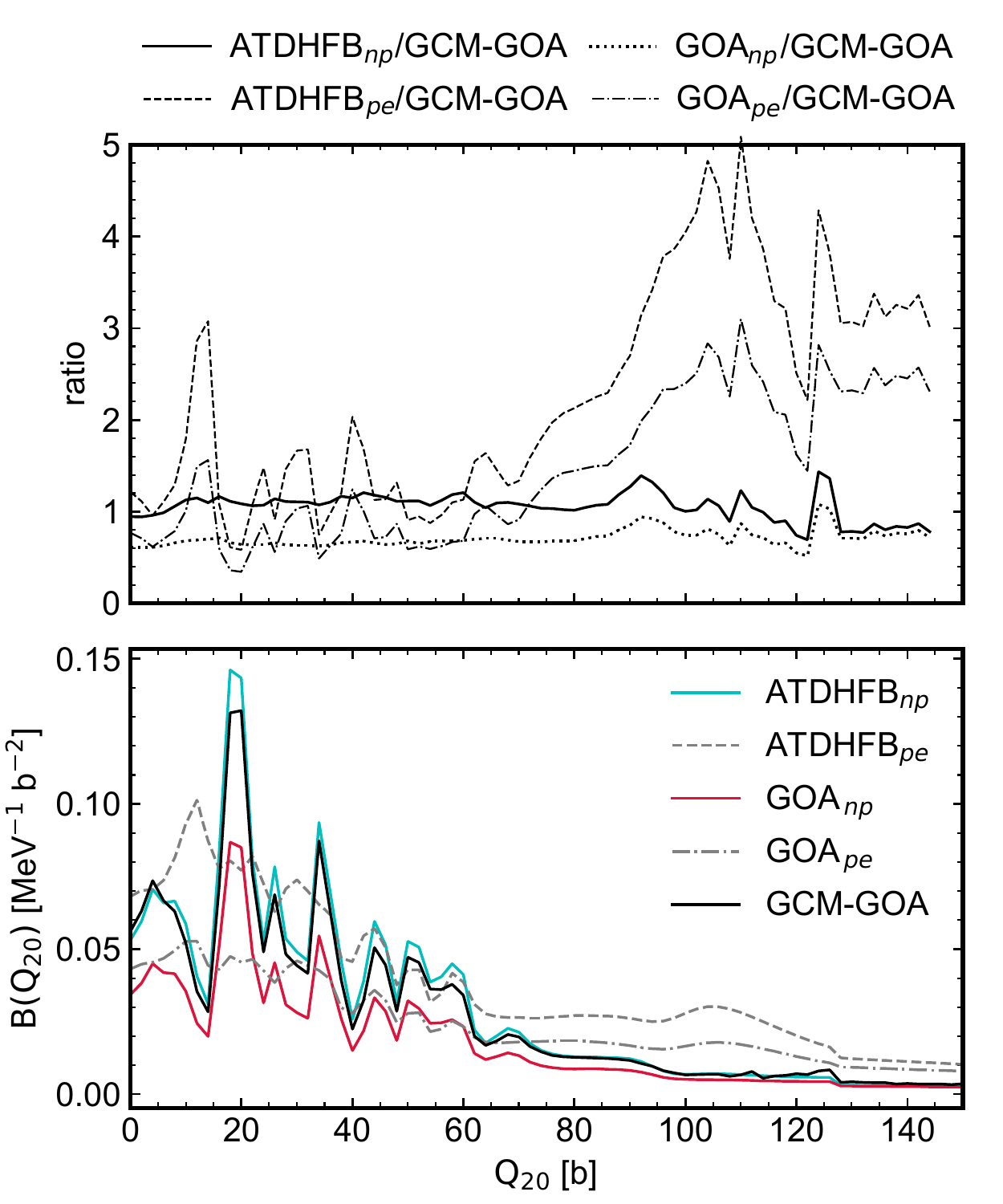}
\caption{\label{fig-2}(Color online) Upper panel: Ratios of different computed
inertias, as a function of the quadrupole moment and for the nucleus $^{234}$U.
Lower panel: The values of the different inertias as a function of the
quadrupole moment.}
\end{figure}
% -----------------------------------------------------------------------------------

The agreement between the non-perturbative inertias diminishes the
uncertainties arising from the ambiguity in the choice of the theoretical scheme
(ATDHFB vs GCM-GOA), but still the suitability of this numerical approximation has
to be proved. In order to address this point we computed the exact GCM-GOA collective
inertias (GCM-GOA) and compared the results with the perturbative and
non-perturbative calculations.  The lower panel of figure~\ref{fig-2} shows the
different inertias computed in this work and the upper panel represents the ratio
of the perturbative and non-perturbative calculations to the exact
GCM-GOA collective inertias.  Surprisingly the GCM-GOA inertias have the same
peak structure and evolution with quadrupole deformation of the non-perturbative
calculations, with a ATDHFB$_{np}$/GCM-GOA ratio very close to one (or
equivalently GCM-GOA$_{np}$/GCM-GOA$\sim0.6$) and virtually independent of
$Q_{20}$. This result brings consistency to the nuclear fission theory,
indicating that the  GCM-GOA inertia can be obtained either by using the
ATDHFB$_{np}$ scheme or the GCM-GOA$_{np}$ mass multiplied by a constant factor around
1.5.
On the other hand, the upper panel of figure~\ref{fig-2} shows that the ratio of
perturbative cranking  to GCM-GOA masses depends on the quadrupole
deformation, with discrepancies in some cases as large as a factor of 5. This
comparison confirms the results found in the non-perturbative study
and indicating the inadequacy of multiplying the perturbative inertias by a phenomenological
factor to grasp the structure of the exact GCM-GOA collective inertia
\cite{Libert10}.

\section{Conclusion}

In summary, this work provides a solution to the uncertainties arising from the
ambiguity in the choice of both the theoretical framework and the numerical
approximations involved in the calculation of collective inertias. Taking
$^{234}$U  as a benchmark, the non-perturbative cranking and exact collective inertias
were calculated for the first time using the Generator Coordinate Method (GCM)
in the Gaussian overlap approximation (GOA)
and compared with the Adiabatic Time Dependent Hartree Fock
Bogolyubov (ATDHFB) non-perturbative and perturbative cranking inertias.

The ATDHFB$_{np}$, GCM-GOA$_{np}$ and
GCM-GOA  inertias present the same peak structure along the whole fission
path, being the GCM-GOA$_{np}$ calculations smaller by a roughly constant factor around
1.5. These inertias show a much richer structure compared to the perturbative
calculations indicating a stronger sensitivity to level crossings.
These results are not only important for fission but also for approximate models
used to describe collective dynamics within the Bohr Hamiltonian or the Collective
Schr\"odinger equation. The use of the non-perturbative inertias along with a phenomenological 
stretching factor can be a good substitute for the more elaborated beyond mean field
calculations with the GCM-GOA.
It would be highly desirable to extend this comparison to the exact ATDHFB
inertia, but the complications arising from the calculation of the inverse of
the linear response matrix $\mathbb{L}$ prevent this comparison for the moment.
Work aimed to compute exact ATDHFB inertias is under way and will be reported
elsewhere.

\section{Acknowledgement} 
We are grateful to G. Martinez-Pinedo for his continuous encouragement in
all the stages of this work as well as for many suggestions and improvements.
SAG  acknowledges support from the U.S. Department of Energy under Award Number
DOE-DE-NA0002847 (NNSA, the Stewardship Science Academic Alliances program), the
Deutsche Forschungsgemeinschaft through contract SFB 1245, and the
BMBF-Verbundforschungsprojekt number 05P15RDFN1. The work of LMR has been
supported in part by Spanish grant Nos FIS2015-63770 MINECO and FPA2015-65929
MINECO.

% --------------------------------------------------------------------------------------------
%                                                                 Appendix
% --------------------------------------------------------------------------------------------

\appendix

% --------------------------------------------------------------------------------------------
%                                                                 Derivatives of the density
% --------------------------------------------------------------------------------------------
\section{Derivatives of the density matrix}

The relationship between the momentum operator matrix elements and
the derivative of the densities with respect to the constraints is
established. Let us consider the density $\rho_{ij}(q) $ 
%\begin{equation}
%\rho_{ij}(q)=\frac{\langle\phi(q)|c_{j}^{\dagger}c_{i}|\phi(q)\rangle}{\langle\phi(q)|\phi(q)\rangle}
%\end{equation}
and the pairing tensor  $\kappa_{ij}(q)$
%\begin{equation}
%\kappa_{ij}(q)=\frac{\langle\phi(q)|c_{j}c_{i}|\phi(q)\rangle}{\langle\phi(q)|\phi(q)\rangle}
%\end{equation}
corresponding to a set of values of the constraints $q$. By shifting
one of the constraints $q_{k}$ by an infinitesimal $\delta q$ we
have to consider
\begin{equation}\label{eq:AppAroqdq}
\rho_{ij}(q_{k}+\delta q)=\frac{\langle\phi(q_{k}+\delta q)|c_{j}^{\dagger}c_{i}|\phi(q_{k}+\delta q)\rangle}{\langle\phi(q_{k}+\delta q)|\phi(q_{k}+\delta q)\rangle} \,.
\end{equation}
%with 
%\begin{align}
%|\phi(q_{k}+\delta q)\rangle & = 
%|\phi(q)\rangle+\delta q\frac{\partial}{\partial q_{k}}|\phi(q)\rangle+O(\delta q^{2}) \\ \nonumber
%%
%& = |\phi(q)\rangle+
%\delta q\sum_{\mu\nu}(Z_{k})_{\mu\nu}\beta_{\mu}^{\dagger}\beta_{\nu}^{\dagger}|\phi(q)\rangle+
%O(\delta q^{2}) \,.
%\end{align}
%Inserting this expression in Eq.~\eqref{eq:AppAroqdq} and using the contractions
With Eqs. (\ref{eq:Phiqdq},\ref{eq:Zq}) and the contractions
$\langle\phi(q)|c_{j}^{\dagger}\beta_{\mu}^{\dagger}|\phi(q)\rangle=V_{j\mu}$
and $\langle\phi(q)|c_{j}\beta_{\mu}^{\dagger}|\phi(q)\rangle=U_{j\mu}$
we easily arrive to
\begin{equation}
\frac{\partial\rho_{ij}}{\partial q_{k}}=\left(UZ_kV^{T}-V^{*}Z^{*}_kU^{\dagger}\right)_{ij}
\end{equation}
and
\begin{equation}
\frac{\partial\kappa_{ij}}{\partial q_{k}}=\left(UZ_kU^{T}-V^{*}Z^{*}_kV^{\dagger}\right)_{ij} \,.
\end{equation}
In order to solve for $Z_k$ in the above expressions it is far more
convenient to work with the unitary block matrix $W$ of Bogoliubov
amplitudes
\begin{equation}
W=\left(\begin{array}{cc}
U & V^{*}\\
V & U^{*}
\end{array}\right)
\end{equation}
and the associated generalized density matrix
\begin{equation}
\mathcal{R}=\left(\begin{array}{cc}
\rho & \kappa\\
-\kappa^{*} & 1-\rho^{*}
\end{array}\right)\,.
\end{equation}
Using them we can write the expressions for the derivatives in a compact
way as 
\begin{equation}
\frac{\partial\mathcal{R}}{\partial q_{k}}=W\left(\begin{array}{cc}
0 & Z_{k} \\
-Z_{k}^{*} & 0
\end{array}\right)W^{\dagger} \,,
\end{equation}
that can straightforwardly be solved for $Z$ 
\begin{equation}
\left(\begin{array}{cc}
0 & Z_{k}\\
-Z_{k}^{*} & 0
\end{array}\right)=W^{\dagger}\frac{\partial\mathcal{R}}{\partial q_{k}}W \,,
\end{equation}
leading to 
\begin{equation} \label{eq:ZpdR}
Z_{k}=U^{\dagger}\frac{\partial\rho}{\partial q_{k}}V^{*}+U^{\dagger}\frac{\partial\kappa}{\partial q_{k}}U^{*}-V^{\dagger}\frac{\partial\kappa^{*}}{\partial q_{k}}V^{*}-V^{\dagger}\frac{\partial\rho^{*}}{\partial q_{k}}U^{*} \,.
\end{equation}

\bibliographystyle{elsarticle-num}
\bibliography{NPinertias}

\end{document}